\documentclass[epj]{svjour}
\usepackage{epstopdf}
\usepackage{graphics}
\begin{document}
\title{Diffusion or bounce back in relativistic heavy-ion collisions?}
\author{Georg Wolschin\inst{1} , Minoru
Biyajima\inst{2} \and Takuya Mizoguchi\inst{3}\vspace{.4cm} 
} 
%\institute{Insert the first address here \and the second here}
\institute{  
  \inst{1}                   
 Institut f{\"ur} Theoretische 
Physik,
% der Universit{\"a}t Heidelberg, 
        Philosophenweg 16,  
        D-69120 Heidelberg, Germany\\
          \inst{2} 
          School of General Education, Shinshu University, Matsumoto 390-8621,
Japan\\
     \inst{3} 
     Toba National College of Maritime Technology, Toba 517-8501, Japan
}
\date{Received: date / Revised version: date}
\abstract{
The time evolution of pseudorapidity distributions of
produced charged hadrons in  d+Au collisions
at  $\sqrt{s_{NN}}$ = 200 GeV is investigated.
Results of
a nonequilibrium-statistical Relativistic Diffusion Model
with three sources are compared with a macroscopic
"bounce back" model that does not allow for statistical equilibration
at large times, but instead leads to motion reversal. 
When compared to the data, the results of the diffusion approach are
more precise, thus emphasizing that the system is
observed to be on its way to thermal equilibrium.
 % end of PACS codes
} %end of abstract
\maketitle
\section{Introduction}
It has recently been shown that rapidity distributions of net protons
\cite{wol03,wol06} and pseudorapidity distributions
of produced charged hadrons \cite{biya04,wols06,kw07} in relativistic heavy-ion collisions can
be described very precisely in a nonequilibrium-statistical
Relativistic Diffusion Model (RDM) with three sources.
There are two sources with initial rapidities close to the beam values that arise mainly
from the valence quarks,
and a third midrapidity source that is mostly due to gluon-gluon collisions,
which reaches statistical equilibrium with respect to the variable pseudorapidity
during the time evolution in the diffusion model.
The particles in this equilibrium source move collectively with very large velocities
\cite{wol06} close
to light velocity similar to a blast wave \cite{sira79}.
They may have hadronized from an equilibrated
quark-gluon plasma.

In case of the asymmetric system d+Au at the highest RHIC
energy $\sqrt{s_{NN}}$ = 200 GeV, we could demonstrate that this
analytical model of a many-particle system on its way to statistical equilibrium
reproduces the complicated features of the experimental distribution
functions to a high degree of accuray. Signatures such as the steeper
slope of the distributions in the deuteron direction as compared
to the gold direction, the
gradual change of the pseudorapidity distributions with centrality,
and the dominance of  particle production in the Au-like rapidity region
towards more central collisions arises in a natural way from the
nonequilibrium-statistical description. The measured pseudorapidity distributions
remain rather asymmetric because strong interaction stops before
the system reaches statistical equilibrium.\\\\
%comments for referee 1

In this work we compare the time evolution of the pseudorapidity
distribution in the nonequilibrium-statistical RDM \cite{wol03,wol06,biya04,wols06}
with a schematic
macroscopic model 
that does $\it{not}$ incorporate statistical equilibrium 
as a limit for large times (except for the midrapidity source),
but instead leads to a separation of the beam-like
distribution functions in pseudorapidity space.
In the course of their time evolution, the mean values of the beam-like
distribution functions change their signs. This corresponds to motion
reversal, or a "bounce back" of the produced particles in the
beam-like sources with respect to the beam directions. Such a model
%reinforces the behaviour of the particles produced in the
%expanding midrapidity source, which are moving
%towards larger absolute  rapidities. It 
does not lead to statistical
equilibrium in the system for large times since the partial distributions separate
in pseudorapidity space.
% for time to infinity. 

In case of an asymmetric system
such as d+Au investigated here, it is possible to decide whether
agreement with the data occurs before or after motion reversal. 
Although a corresponding distinction is not possible for symmetric
systems such as Au+Au, the results for d+Au will indicate that
symmetric systems behave in an analogous way. 

The Relativistic Diffusion Model is reconsidered briefly in the
next section. Afterwards, the bounce back approach is outlined,
and both models with their corresponding time evolution are then compared  
to the available PHOBOS minimum-bias  \cite{bbb04} and centrality dependent
 \cite{bbb05}
d+Au data at  $\sqrt{s_{NN}}$ = 200 GeV, and to each other.
Such a comparison is conceptually quite important because a preference
for the diffusion case would emphasize that the relativistic systems
are indeed observed to be on their way to thermal equilibrium.
\newpage
\section{Nonequilibrium-Statistical Approach}
The nonequilibrium-statistical approach is based on diffusion equations
which eventually lead to statistical equilibrium for large times.
In the three-sources model, we have successfully used linear Fokker-Planck
equations (FPEs)  \cite{wol03,biya04,wols06} for the components $R_{k}(y,t)$
of the distribution function for produced charged hadrons
in rapidity space
\begin{equation}
\frac{\partial}{\partial t}R_{k}(y,t)=
\frac{1}{\tau_{y}}\frac{\partial}
{\partial y}\Bigl[(y-y_{eq})\cdot R_{k}(y,t)\Bigr]
+\frac{\partial^2}{\partial y^2}\Bigl[D_{y}^{k}
\cdot R_{k}(y,t)\Bigr]
\label{fpe}
\end{equation}\\
with the rapidity $y=0.5\cdot \ln((E+p)/(E-p))$.
The diagonal components $D_{y}^{k}$ of the rapidity
diffusion tensor contain the
microscopic physics in the respective beam-like $(k=1,2)$
and central $(k=3)$ regions. They account for the broadening of the
distribution functions through interactions and particle creations.
The off-diagonal terms of the diffusion tensor account for 
correlations between the three sources. They are expected to be small,
and we neglect them in this work.
The rapidity relaxation time $\tau_{y}$ determines the speed of the 
statistical equilibration in $y$-space.

Whereas an equilibrium-statistical description of relativistic 
heavy-ion collisions works well for certain observables 
such as abundance ratios of produced hadrons (Òthermal modelÓ), 
a nonequilibrium-statistical formulation as outlined here
is required for distribution functions and other more 
sophisticated observables. 
A description of the dynamics 
should be part of such a formulation. In the 
present work, the dynamics enters through the 
time variable in the FPE that describes the evolution 
of the three sources for particle production in 
rapidity space. 

For time to infinity, both the mean values 
and the variances of the three sources reach the 
equilibrium values with respect to the variable rapidity. 
Then the incoherent sum of the three distributions represents 
the overall statistical equilibrium distribution in 
rapidity space Ð and after Jacobi-transformation, in 
pseudorapidity space. Comparison with the data for 
d+Au at the highest RHIC energy shows, however, 
that strong interaction stops before this equilibrium distribution is reached.

In the nonequilibrium-statistical model as well as
in the bounce back approach, the initial particle production in the three
sources is assumed to occur with full strength at very
short times. In the diffusion case, we use in this work the initial conditions
$R_{1,2}(y,t=0)=\delta(y\pm y_{max})$
with the maximum rapidity $y_{max}=5.36$ at
the highest RHIC energy of $\sqrt{s_{NN}}$ = 200 GeV
(beam rapidities are $y_{1,2}=\mp y_{max}$), and $R_{3}(y,t=0)=\delta(y).$
%with the equilibrium value of the rapidity $y_{eq}$. 
A midrapidity gluon-dominated symmetric source had also been proposed by
Bialas and Czyz \cite{bc05}. It may originate from a thermally equilibrated
quark-gluon plasma.

These initial conditions are slightly modified for the central source as 
compared to our previous investigation \cite{wols06}, where we used
$R_{3}(y,t=0)=\delta(y-y_{eq})$. The motivation is twofold: We want to 
use identical initial conditions when comparing diffusion and 
bounce back approach, and we want to investigate whether the modified
initial condition allows to maintain the quality of the agreement with
the data in the diffusion approach.

The new initial condition for the midrapidity source corresponds to 
initial (t=0) particle production at rest, independently of the mass
of the collision partners: the third source is created at y=0. Hence, this
gluon-dominated source is initially insensitive to the mass distribution in
the system, and it is only at larger times - when the particles in
this source have already been created - that the drift towards the equilibrium
value sets in. Since there are very few valence quarks in the
midrapidity region at the highest RHIC energy, this initial condition
is probably more realistic than the one we used in \cite{wols06}.

The mean values in the three sources have the time dependence
\begin{equation}
<y_{d1,2}(t)>=y_{eq}[1-\exp(-t/\tau_{y})] \mp y_{max}\exp{(-t/\tau_{y})}
\label{mean}
\end{equation}
for the sources (1) and (2), and 
\begin{equation}
<y_{d3}(t)>=y_{eq}[1-\exp(-t/\tau_{y})]
\label{mean3}
\end{equation}
for the moving central source. The three mean values reach the equilibrium
value for time to infinity. In our previous RDM-calculation \cite{wols06}
with slightly different initial condition, 
the mean value of the central source was at the equilibrium value 
$<y_{3}(t)>=y_{eq}$ independently of time, thus assuming instant
equilibration in this source regarding the mean values.
The variances are obtained as in \cite{wb06}.
It will turn out that for d+Au at the highest RHIC energy, we can not
determine from a comparison with the data which of the two possibilities
for the initial conditions of the central source is more realistic because
the $\chi^{2}$ is nearly identical in both cases.

We follow the subsequent diffusion-model time evolution
in pseudorapidity space up to the interaction time $\tau_{int}$,
when the produced charged hadrons cease to interact
strongly. The quotient $\tau_{int}/\tau_{y}$ is determined from the minimum $\chi^{2}$
with respect to the data, simultaneously with the minimization
of the other free parameters - in particular, the
variances of the three partial distribution functions,
and the number of particles produced in the central source.
In the nonequilibrium-statistical approach,
the equilibrium value of the rapidity and its dependence on
centrality is calculated from energy and momentum conservation
in the system of participants as
\begin{eqnarray}
y_{eq}(b)=\qquad\qquad\qquad\qquad\qquad\qquad\qquad\qquad\qquad
\nonumber\\
\frac{1}{2}\ln\frac{<m_{1}^{T}(b)>\exp(-y_{max})+<m_{2}^{T}(b)>
\exp(y_{max})}
{<m_{2}^{T}(b)>\exp(-y_{max})+<m_{1}^{T}(b)>\exp(y_{max})}
\label{yeq}
\end{eqnarray}\\
with
% the beam rapidities y$_{b} = \mp y_{max}$, 
the transverse masses $<m_{1,2}^{T}(b)>=
\sqrt(m_{1,2}^2(b)+
\newline
<p_{T}>^2)$, and masses
m$_{1,2}(b)$ of the "target"(Au)- and "projectile"(d)-participants 
that depend on the impact parameter $b$. The average 
numbers of participants 
\newline
$<N_{1,2}(b)>$ from the Glauber
calculations reported in \cite{bbb05} for minimum bias d + Au 
at the highest RHIC energy are
%\newline
 $<N_{1}>=$6.6, $<N_{2}>=$1.7,
which we had also used in \cite{wols06,wb06}.
With $<p_{T}>= 0.4$ GeV/c the result is $y_{eq}=-0.664$,
with $<p_{T}>= 1 $ GeV/c we obtain $y_{eq}=-0.60$.

The average numbers of charged particles in
the target- and projectile-like regions $N_{ch}^{1,2}$ are 
proportional to the respective
numbers of participants $N_{1,2}$,
\begin{equation}
N_{ch}^{1,2}=N_{1,2}\frac{(N_{ch}^{tot}-N_{ch}^{eq})}{(N_{1}+N_{2})}
\label{nch}
\end{equation}
with the constraint $N_{ch}^{tot}$ = $N_{ch}^1$ + $N_{ch}^{2}$ +
$N_{ch}^{eq}$.
Here the total number of charged particles 
$N_{ch}^{tot}$ is determined from the data. The average number
of charged particles in the equilibrium source $N_{ch}^{eq}$ is a
free parameter that is optimized together with the variances
and $\tau_{int}/\tau_{y}$ in a $\chi^{2}$-fit of the data
using the CERN minuit-code. Due to the accuracy of the data, 
these five free parameters (for both scenarios, diffusion and bounce back)
are determined with great precision.

\begin{figure*}
\resizebox{0.9\textwidth}{!}{%
 \includegraphics{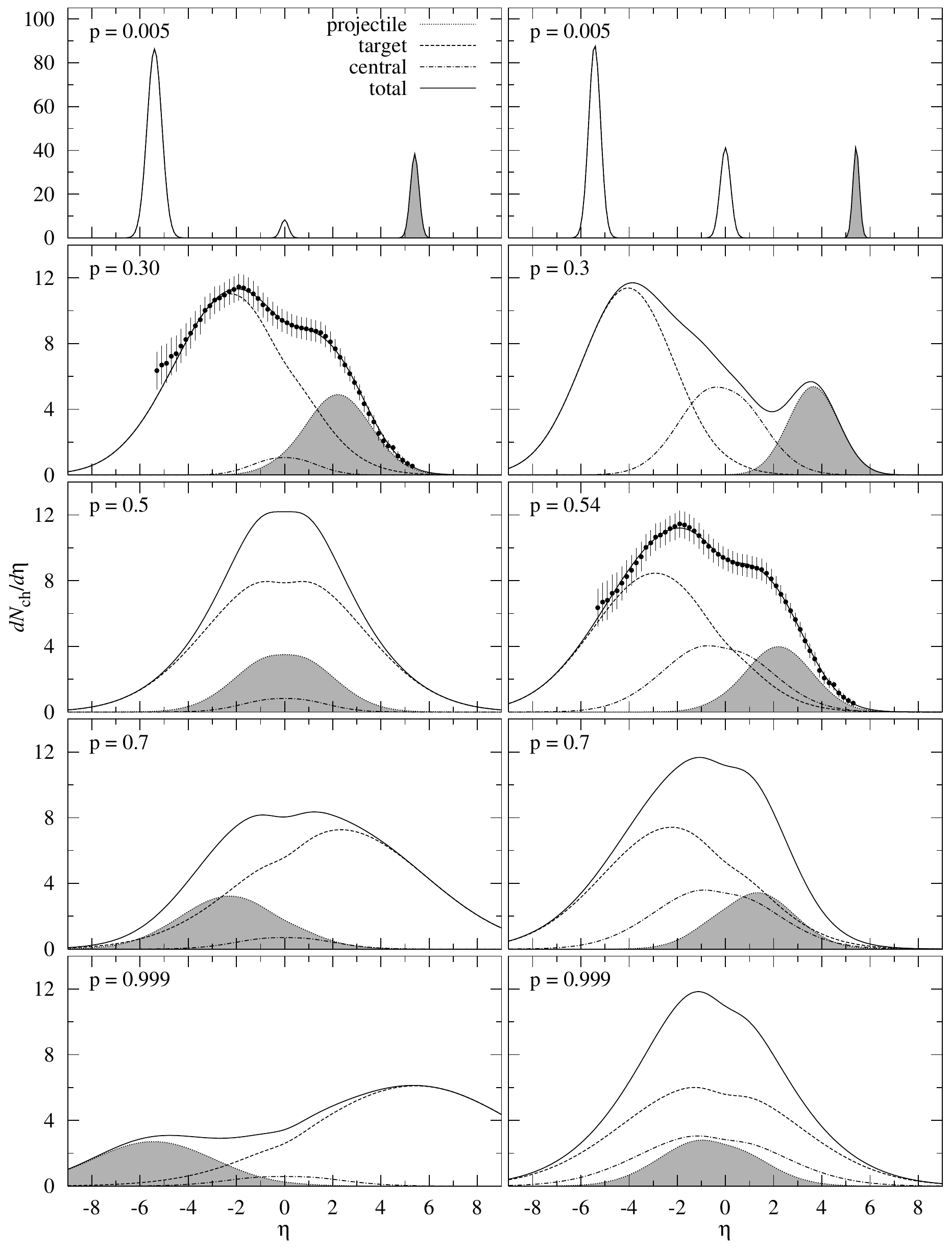}}
 %\includegraphics{fig1.eps}}
%\vspace*{18cm}       % Give the correct figure height in cm
%\end{document}
%\newpage
\caption{Time evolution of pseudorapidity distributions for
produced charged particles from minimum-bias
%\newline
 d + Au collisions at $\sqrt{s_{NN}}$ = 200 GeV. Results for five
time-steps ($p$-values, cf. text) are shown. In the left-hand column,
the bounce back calculation with three sources is displayed.
Here the Au- and d-like (grey) distributions reverse their motion in the mean
at large times, and then tend to separate.
Agreement with the data 
can only be achieved "on the way in" at $p$=0.30, before bounce back occurs. 
Dash-dotted curves show
the midrapidity sources for hadron production.
The right-hand column shows the diffusion-model calculation
where statistical equilibrium would be achieved for large times
at the equilibrium value $\eta_{eq}$. The diffusion-model evolution reaches
agreement with the data much slower, at p=0.54.
The minimum  $\chi^{2}/d.o.f.$-values with respect to the
PHOBOS data \cite{bbb04} are 2.4/48 for diffusion, and 7.4/48 for bounce back.} 
\label{fig1}
%\end{center}
\end{figure*}

The FPE is solved analytically as outlined in \cite{wols06,wb06},
the result is converted to pseudorapidity space, and compared to data,
fig.\ref{fig1}, right-hand column. Here the time evolution parameter $p$
in the numerical calculation is defined as
\footnote{
There is a difference of 
a factor of two in the exponent as compared to the definition of $p$ used in 
\cite{wb06}, which causes different $t/\tau_{y}$ values for given $p$.}
\begin{equation}
p=1-\exp(-t/\tau_{y}) .
\label{pe}
\end{equation}
%As compared to the calculations shown in \cite{wols06,wb06},
%the $p$-value of the time evolution that yields minimum
%$\chi^{2}$ with respect to the PHOBOS
%data \cite{bbb04} and also the other four free parameters have 
%slightly different values ($p$=0.53 in this work as compared to $p$=0.54
%in \cite{wb06}, different values for the three variances, and the number
%of particles in the central source, see table 1). This is mostly due to the fact that
%the numbers of participants $<N_{1,2}>$ have been modified 
%from the Glauber results in order to further improve the agreement with the data.

We have shown in \cite{wols06,wb06} that with this RDM approach, the 
centrality dependence of the measured pseudorapidity 
distributions \cite{bbb05} from
central to very peripheral collisions can be modeled in
considerable detail. For peripheral collisions, the asymmetry
of the overall distribution is not yet pronounced because here the
d- and the Au-like partial distributions are similar in size due
to the small number of participants. Towards more central collisions,
the number of gold participants rises, and the corresponding
partial distribution of produced particles becomes more important.
In addition, the distributions drift towards the equilibrium value.
Both effects produce the asymmetric shape, which is also 
seen in minimum-bias.

The minimum-bias result that we present here in more detail also
shows the asymmetric shape, which is very well reproduced
in the diffusion calculation. At larger values of the time evolution
parameter $p$, all three subdistributions tend to become symmetric
in $y$ with respect to the equilibrium value $y_{eq}$, indicating the approach
to thermal equilibrium. At $p$=0.999, the equilibrium state is already
closely approached. The slight asymmetry is due to the conversion from
rapidity- to pseudorapidity space which tends to produce a dip at
$\eta=0$. For time to infinity,
statistical equilibrium in pseudorapidity space would be reached.
%\newpage
%\end{document}
%\end{flushleft}
\begin{table}
\caption{Produced charged hadrons 
 in minimum-bias d + Au  
collisions at $\sqrt{s_{NN}}$ =
200 GeV, y$_{1,2}=\mp$ 5.36 in the "bounce back" model
(1st line),
and in the Relativistic Diffusion Model (2nd line).
%The ratio $\tau_{int}/\tau_{{y}}$ determines how fast the system 
%of produced charged particles equilibrates in rapidity space.
%The effective diffusion coefficient in the central source is $D_{y}^{eff}$.
The equilibrium value of the rapidity in the RDM
is $y_{eq}$, the time parameter (see text) is $p$,
the corresponding value of interaction time over relaxation
time is $\tau_{int}/\tau_{y}$, the variance of the central source in
$y-$space is $\sigma_{3}^{2}$. The number of produced charged particles 
is $N_{ch}^{1,2}$ for the sources 1 and 2 and $N_{ch}^{3}$ for the central
source, the percentage of
charged particles produced in the midrapidity source is $n_{ch}^{3}$.}
\vspace{.1cm}
\label{tab1}
%\begin{center}
\begin{tabular}{cccccccc}
\hline
$y_{eq}$&$p$&$\tau_{int}/\tau_{y}$&
$\sigma_{3}^{2}$&$N_{ch}^{1}$&$N_{ch}^{2}$&$N_{ch}^{3}$&$n_{ch}^{3}$(\%)\cr\\
\hline
--&0.30&0.36& 1.32  &   66&  17&3&3\cr
-0.664&0.54&0.78&4.19 &  55 & 14& 22 & 24\cr 
\hline
\end{tabular}
%\end{center}
\end{table}

It is interesting to determine actual values for the 
diagonal components of the diffusion tensor from  the 
expression for the variances, eq.(\ref{var}), provided the interaction 
time is known. As an example, for an interaction time of 7fm/c in d+Au
and $\tau_{int}/\tau_{y}$=0.78, the resulting values of the 
three diffusion coefficients in minimum-bias are $D_{y}^{1,2,3}$
= 1.55, 0.46 and 0.98 c/fm. These are effective 
values as explained in ref.\cite{wol06} because they include the 
influence of collective expansion in addition to the statistical fluctuations. 

\section{Bounce Back Approach}
In this schematic model of motion reversal in pseudorapidity space,
the beam-like partial distribution functions change their signs
in the course of the time evolution of the collision. The 
 behaviour of the system is approximated through a linear
partial differential equation which is similar to a Fokker-Planck
equation, but differs in that it does not lead to statistical
equilibrium in the system for large times. Instead it causes
a separation of the beam-like sources in rapidity space:
%Here bb-FPE
\begin{equation}
\frac{\partial}{\partial t}R_{k}(y,t)=
\frac{1}{\tau_{y}}\frac{\partial}
{\partial y}\Bigl[(y-y_{k})\cdot R_{k}(y,t)\Bigr]
+\frac{\partial^2}{\partial y^2}\Bigl[D_{y}^{k}
\cdot R_{k}(y,t)\Bigr].
\label{bbe}
\end{equation}\\
Here the drift term does not account for the 
equilibration towards $y_{eq}(b)$ as in the FPE,
but rather describes the motion of both peripheral distributions
in rapidity space towards smaller absolute values of the rapidity, and
eventually into the other hemisphere of rapidity space - which
means motion reversal, or bounce back. The drift term 
that induces this behaviour of the beam-like partial
distributions 
is $(y-y_{k})=(y\mp y_{max})$ for  k=1,2.
The drift term for the midrapidity source is
 $(y-y_{3})=y$.
 
We use the same $\tau_{y}$ as in the RDM case in order to 
have a direct comparison of the two scenarios. 
In the Ôbounce-backÕ scenario, there is no 
relaxation towards statistical equilibrium and hence, 
$\tau_{y}$ should not be considered to be a relaxation time 
in this case, it is just a time parameter that controls the 
speed of the movement in y- (or $\eta$-)space.

Because
the three-component system does not approach 
statistical equilibrium with respect to the variable
rapidity for time to infinity, 
the underlying system of partial differential equations
for the bounce back case differs in an important aspect from the ordinary
Fokker-Planck, or Uhlenbeck-Ornstein \cite{uhl30}
framework, although it looks formally quite similar
\footnote{A preliminary version of the bounce back approach had been proposed
in \cite{bm04}}. 

The initial conditions for the beam-like partial distribution
functions are taken to be the same as in the nonequilibrium-statistical 
diffusion approach.
Since the \\equation is linear, a superposition of the partial distribution
functions  using the initial conditions
$R_{1,2}(y,t=0)=\delta(y\pm y_{max})$
with the maximum rapidity $y_{max}$, and $R_{3}(y,t=0)=\delta(y)$  
yields the exact solution. In the solution, the mean values 
are obtained analytically from the moments 
equations as
\begin{equation}
<y_{bb1,2}(t)>=\pm y_{max}[1-\exp(-t/\tau_{y})] \mp y_{max}\exp{(-t/\tau_{y})}
\label{meanbb}
\end{equation}
for the sources (1) and (2), and 
\begin{equation}
<y_{bb3}(t)>=0
\label{meanbb1}
\end{equation}
for the third source, which rests at 0 in the bounce back case during the
whole time evolution.
It is therefore obvious that the mean values of the beam-like
distribution functions start their time
evolution at t=0 with values $y_{1,2}(t=0)=\mp y_{max}$, and for 
time to infinity they reach $y_{1,2}(t\rightarrow \infty)=\pm y_{max}$.
Hence, they change from one rapidity hemisphere into the other
one at  a value of the bounce back time
\begin{equation}
\frac{t_{bb}}{\tau_{y}}=-\ln(1/2)=0.6932
\label{tbb}
\end{equation}
which corresponds to a value of the time evolution parameter
p=0.5, as can be seen in the middle frame of the bounce back case in fig. 1. 
We use the same $\tau_{y}$ in sections 3,4 in order to have 
a direct comparison of the two scenarios. 
Of course in the bounce back scenario, there is no relaxation 
towards statistical equilibrium and hence, $\tau_{y}$ should 
not be considered to be a relaxation time in this case. Here it
 is a time parameter that controls the speed of the movement in 
 y- (or $\eta$-)space.

The variances are as in the diffusion case
\begin{equation}
\sigma_{1,2,3}^{2}(t)=D_{y}^{1,2,3}\tau_{y}[1-\exp(-2t/\tau_{y})].
\label{var}
\end{equation}

The conversion to pseudorapidity is required
because particle identification is not available:
$\eta=-$ln[tan($\theta / 2)]$ with the scattering angle $\theta$
measured relative to the direction of the deuteron beam. Hence,
particles that move in the direction of the gold beam have negative, particles 
that move in the deuteron direction have positive
pseudorapidities.
The conversion from $y-$ to $\eta-$
space of the rapidity density
\begin{equation}
\frac{dN}{d\eta}=\frac{p}{E}\frac{dN}{dy}=
J(\eta,\langle m\rangle/\langle p_{T}\rangle)\frac{dN}{dy} 
\label{deta}
\end{equation}
is performed through the Jacobian
\begin{eqnarray}
\lefteqn{J(\eta,\langle m\rangle/\langle p_{T}\rangle) = \cosh({\eta})\cdot }
\nonumber\\&&
\qquad\qquad[1+(\langle m\rangle/\langle p_{T}\rangle)^{2}
+\sinh^{2}(\eta)]^{-1/2}.
\label{jac}
\end{eqnarray}We approximate the average mass $<m>$ of produced charged hadrons in the
central region by a value somewhat higher than the pion mass $m_{\pi}$, and use a
mean transverse momentum $<p_{T}>$ = 0.4
GeV/c such that the relevant parameter in the Jacobian becomes
$<m>/<p_{T}>=0.45/c$.

  Due to the conversion from $y-$ to $\eta-$space,
the partial distribution functions are different from Gaussians.
The charged-particle distribution in rapidity space is obtained
in both the diffusion model, and the bounce back case as incoherent 
superposition of nonequilibrium and local equilibrium solutions of
 (\ref{fpe}) 
\begin{eqnarray}
    \lefteqn{
\frac{dN_{ch}(y,t=\tau_{int})}{dy}=N_{ch}^{1}R_{1}(y,\tau_{int})}\nonumber\\&&
\qquad\qquad +N_{ch}^{2}R_{2}(y,\tau_{int})
+N_{ch}^{3}R_{3}(y,\tau_{int})
\label{normloc1}
\end{eqnarray}
with the interaction time $\tau_{int}$ (total integration time of the
differential equation). In the present work, the integration is 
stopped at the value of $\tau_{int}/\tau_{y}$ that produces the
minimum $\chi^{2}$ with respect to the data and hence, the
explicit value of $\tau_{int}$ is not needed as an input. 
%The result for central collisions is $\tau_{int}/\tau_{y} 
%\simeq 0.4$, other values are given in Table \ref{tab1}.
The resulting values for $\tau_{int}/\tau_{y}$ are given
in table \ref{tab1} together with the widths of the central
 distributions, and the particle numbers in the three sources.
 
 We show the time evolution in the bounce back case together with
 the fit to the PHOBOS data \cite{bbb04} in the left-hand column of 
 fig.\ref{fig1}. It is evident that the two beam-like distribution
 functions move towards smaller pseudorapidities as time
 increases, reach agreement with the data at $p$=0.30 before motion reversal
 in the mean is achieved, then move on in pseudorapidity
 space to reverse their motion in the mean at $p$=0.5, and finally proceed
 until they would separate according to the time evolution given
 by eq.(\ref{bbe}). At $p=0.999$, a very broad distribution 
 in pseudorapidity has emerged, which deviates considerably
 from the near-thermal distribution that can be seen in
 the diffusion case at the same time step.
 
  The RHIC data of the asymmetric d+Au system
  show very clearly that motion reversal in the mean is not observed,
  since the maximum occurs at negative pseudorapidities
  in the gold direction. Only a certain fraction of the produced particles
  moves opposite to the respective beam direction. 
 Although in the bounce back case,  the underlying
 system of differential equations shows a memory in the
exit channel (P- and T-like distributions emerge separately at
large times) the effect is not observed, because the measurement occurs
before that would happen.
  
  Comparing to the diffusion-model time evolution on the right-hand side of fig.1,
  it is obvious that agreement with the data is reached much faster in the
  bounce back case ($p$=0.30 as compared to $p$=0.54). The reason is
  found in the respective equations for the mean values (\ref{mean}), (\ref{meanbb}).
  To first order in $t/\tau_{int}$, the diffusion result is
 \begin{equation}
<y_{d1,2}>\approx\mp y_{max}\pm y_{max}t/\tau_{y}
\label{dm}
\end{equation}
whereas in the bounce back case
 \begin{equation}
<y_{bb1,2}>\approx\mp y_{max}\pm 2y_{max}t/\tau_{y}
\label{bm}
\end{equation}
and hence, the drift in the bounce back case for fixed $t/\tau_{y}$ is much
stronger than in the diffusion case.

Because the underlying partial differential
 equation conserves the norm, the particle number in each 
 distribution function does not change in our schematic model
 once the particles are created. The $\chi^{2}$-minimization
 yields a different partition of the produced charged hadrons among
 the three sources as compared to the diffusion case, table \ref{tab1}.
 In particular, the percentage of particles in the midrapidity source 
 is much smaller as in the diffusion case, only 3 $\%$ instead of 24 $\%$.
 As is evident from fig.\ref{fig1} and table \ref{tab1}, the diffusion case
 yields better agreement with the data.
 %Here fig.2
  \begin{figure*}
%\begin{center}
\resizebox{0.9\textwidth}{!}{%
 \includegraphics{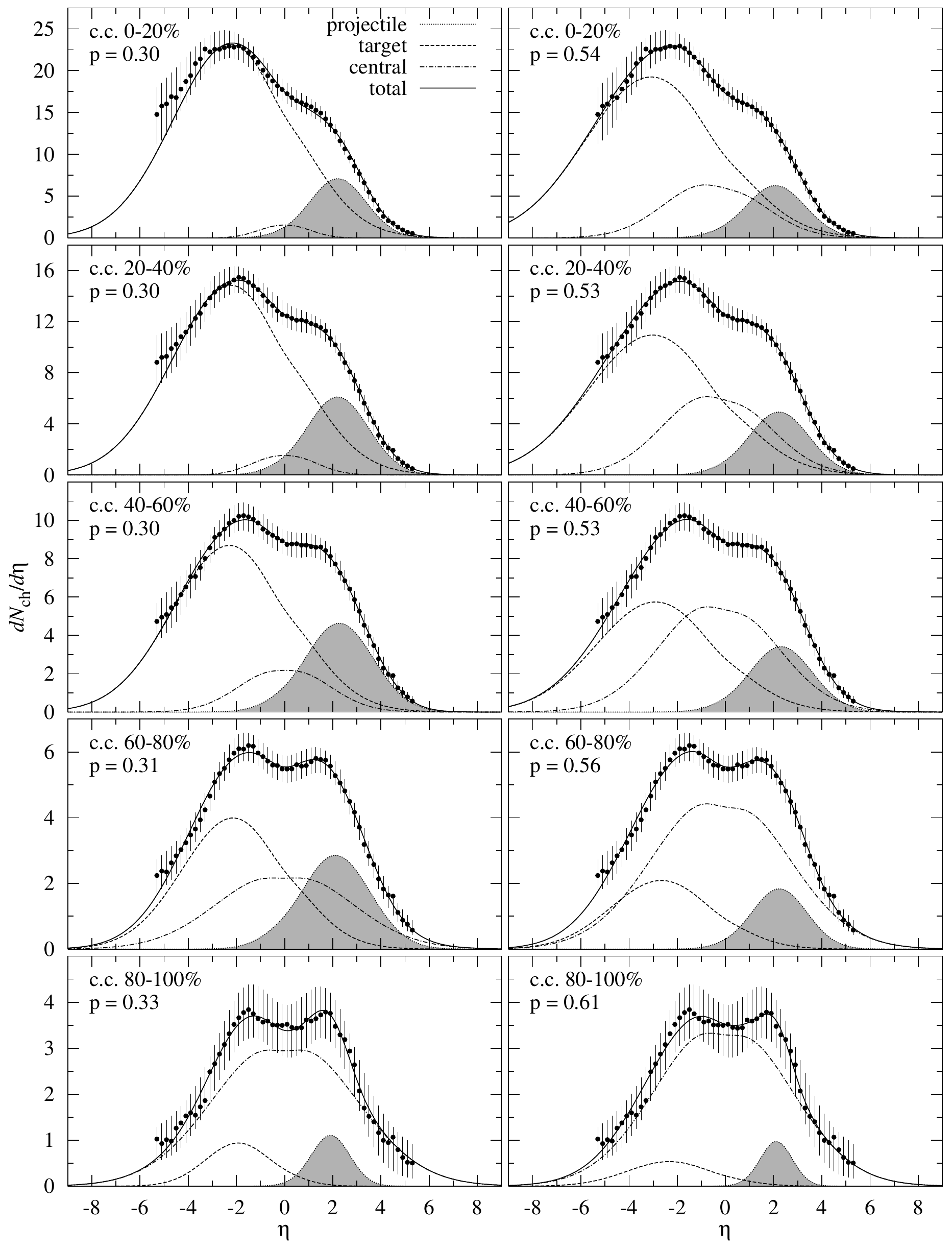}}
\caption{Calculated pseudorapidity distributions of 
produced charged particles in
%\newline
 d + Au collisions at $\sqrt{s_{NN}}$ = 200 GeV for five
different centralities. Central collisions are shown in the top
frames, peripheral at the bottom. In the left-hand column, results
of the bounce back calculation with three sources are shown.
The right-hand column gives results of the diffusion-model calculation
where statistical equilibrium would be achieved for large times
at the equilibrium value $\eta_{eq}$. The diffusion-model evolution
provides a more detailed agreement with the
PHOBOS data \cite{bbb05}.} 
\label{fig2}
%\end{center}
\end{figure*}
However, both the bounce back and the diffusion evolution 
still appear to be acceptable models, and a clear distinction between
 the two becomes apparent only in the subsequent undetected time evolution.
Hence,  it seems not straightforward so far to conclude from this analysis that the 
 system is observed to be on its way to statistical equilibrium. To prove that this
 is likely to be the correct physical interpretation, as outlined in
 \cite{wol03,wol06,biya04,wols06,wb06}, one indeed needs additional
 physical information.
 
 A possibility to obtain this information is the
 investigation of the centrality dependence as performed in
 \cite{wols06,wb06} for the RDM-case.
 %(with slightly different
% initial conditions, see also preceding chapter). 
Such an impact-parameter
 dependent analysis is very sensitive to the details of the model. 
 We have performed this analysis for the bounce back case, and repeated it
 for the RDM-case with the same initial condition for the central source. We
 use the same Glauber values for the average number
 of participants in each centrality bin as in \cite{bbb05,wols06,wb06}.
 
The results as presented in  fig.\ref{fig2} show
that the RDM gives  consistently better agreement with the data for all
 centralities, in particular for more central collisions where the diffusion
 approach is expected to function best because of the larger
 number of participants. The $\chi^{2}-$values of the
 bounce back calculation vary from 12.3 for $0-20\%$ (central) to
 4.9 for $80-100\%$ (peripheral), as compared to 1.5 (central)
 and  4.6 (peripheral) in the diffusion case.
 To obtain better results in the bounce back case,
deviations from the Glauber values would be required
%comment added in response to report 2
which are not realistic.
As an example, for $<N_{1}>=7$ and $<N_{2}>=1$ in minimum bias
we obtain $\chi^{2}=2.1$, which is comparable to the diffusion  
result with the Glauber values $<N_{1}>=6.6$ and $<N_{2}>=1.7$,
 but the value for the deuteron participants is unrealistic.

The interpretation
provided by the Relativistic Diffusion Model
is also favoured by
the experimental fact that the distributions of produced particles in
transverse momentum space are very close to thermal distributions.
Hence, it seems natural to infer that the system tends to  approach 
thermal equilibrium, and that this tendency is seen
in the longitudinal variables as well.

We have confined this investigation to a comparison with
the PHOBOS data because these provide a large number of data
points with high accuracy, giving better constraints of the free
parameters than the corresponding BRAHMS \cite{ars05},
and STAR results \cite{abe07}.
% \newpage
\section{Conclusion}
To conclude, we have compared two schematic models for
charged-hadron production in relativistic  d+Au collisions
at $\sqrt{s_{NN}}$ = 200 GeV for minimum bias,
 and depending on centrality.
The nonequilibrium-statistical Relativistic Diffusion Model (RDM)
describes the gradual approach of the system towards
statistical equilibrium. The rapidity
distribution functions become symmetric at large times with 
respect to the equilibrium value of the rapidity, which is 
obtained from energy- and
momentum conservation in the system of participants.

Since strong interaction stops before the
system reaches equilibrium, the pseudorapidity distribution
for minimum-bias collisions remains rather asymmetric.
In this approach, all the details of the experimental distribution function such as
the different slopes in the respective gold and deuteron directions are
precisely reproduced for all centralities with good $\chi^{2}$-values.
The midrapidity source contains 24${\%}$ of the produced 
charged hadrons. 

In the bounce back approach, the underlying differential equation would
lead to a re-separation of the partial beam-like distribution functions
%in pseudorapidity space, rather than to equilibrium. The 
in $\eta-$ space, rather than to equilibrium. The 
midrapidity source is very small in size (3${\%}$).
%of the produced particles for minimum bias collisions.
%It does not drift towards the equilibrium
%value, but remains centered at rest ($<\eta_{3}(t)>=0$). 
It remains centered at rest, $<\eta_{3}(t)>=0$.

When comparing the bounce back
time evolution to the available PHOBOS-data \cite{bbb04},
it turns out that satisfactory agreement can only be achieved
"on the way in" before motion reversal, because the
experimental distribution peaks in the negative 
pseudorapidity hemisphere, fig.\ref{fig1}. 
The data are reached faster during the time evolution
as compared to the diffusion case because the drift
is stronger. The system 
is quite transparent, and it is not possible to actually
observe motion reversal (bounce back) in the mean of the produced-hadron
distributions at RHIC energies. Most likely, this is true at LHC 
energies as well, since the transparency is expected
to be even more pronounced at higher energies \cite{kw07}.

This result is particularly relevant
for the interpretation of heavy symmetric systems like Au+Au or Pb+Pb,
which are expected to behave similarly, but where the two possibilities
- agreement with the data "on the way in" (partial transparency),
or after motion reversal (bounce back) - can not be distinguished easily due to the 
symmetry of the system. From the analogy with our investigation of an
asymmetric system, however, it appears certain that partial transparency
is the correct physical description at RHIC energies and above.
For symmetric systems with $y_{eq}=0$, the two descriptions outlined in this work 
are difficult to distinguish when compared to data.
%because agreement
%with the data occurs before motion reversal.
% the actual fits also
%give quite similar results.

For asymmetric systems, however, the differences of the two models are quite pronounced,
in particular when the centrality dependence is considered.
The RDM approach is more
precise, and we expect that it yields converging $\chi^{2}-$minimizations with
respect to the experimental pseudorapidity distributions for every
relativistic heavy-ion collision also at other energies,
for example, Si+Al at AGS or
S+Au at  SPS energies. The distribution of produced particles among
the three sources, however, will vary substantially depending on
incident energy, size and asymmetry of the system, and centrality. The third
source vanishes at sufficiently low energy, for example, at 
$\sqrt{s_{NN}}$ = 19.6 GeV for Au+Au \cite{kw07}.
%In other
%asymmetric systems such as Si+Al and S+Au, it is also more difficult
%to distinguish the two cases as compared to the very
%asymmetric d+Au system, where the distinction turns out to be possible.

In summary, consistency with the data at RHIC energies can only be
achieved based on partial transparency without motion 
reversal in the mean, independently of the specific model.
Comparing bounce back and nonequilibrium-statistical approach
in detail, the agreement of the diffusion approach with the
data is definitely better. This result supports the
view that the relativistic many-body system is 
indeed observed to be on its way towards statistical equilibrium.
\section{Acknowledgements}
Two of the authors (MB and TM) would like to thank
RCNP at Osaka University for discussions at meetings.
The work is supported by DFG under contract
No. STA 509/1-1.
\\

\end{document}